\documentstyle[prl,aps,preprint,tighten,floats,aps,epsfig,psfig]{revtex}
\def\btabl{\begin{table}}   \def\etabl{\end{table}}
\def\bea{\begin{eqnarray}}   \def\eea{\end{eqnarray}}
\def\bnn{\begin{eqnarray*}}   \def\enn{\end{eqnarray*}}
\def\beq{\begin{equation}}   \def\eeq{\end{equation}}  
\def\btabu{\begin{tabular}}   \def\etabu{\end{tabular}}
\def\bec{\begin{displaymath}} \def\eec{\end{displaymath}}
\def\nn{\nonumber}
\def\eqref#1{(\ref{#1})}
\renewcommand{\baselinestretch}{1.2}
\begin{document}
\draft
\date{\today}
\preprint{\vbox{\baselineskip=13pt
\rightline{CERN-TH/97-350}
\vskip0.2truecm
\rightline{FAMNSE-97/19}
\vskip0.2truecm
\rightline{LPTHE Orsay-97/65}
\vskip0.2truecm
\rightline{hep-ph/9712266}}}

\title{Multibody neutrino exchange in a neutron star:\\
 neutrino sea and border effects}
\author{As. Abada$^{a}$, 
O. P\`ene$^b$ and J. Rodr\'\i guez-Quintero$^c$ \footnote{
e-mail: abada@mail.cern.ch, jquinter@cica.es, 
pene@qcd.th.u-psud.fr.}} 

\address{{\small 
$^a$ Theory Division, CERN,CH-1211Geneva 23, Switzerland.\\
$^b$ Laboratoire de Physique Th\'eorique et Hautes
Energies\\
Universit\'e de Paris XI, B\^atiment 211, 91405 Orsay Cedex,
France.
\\
$^c$ Departamento de F\' \i sica At\'omica, Molecular y Nuclear, Universidad 
de Sevilla \\ P.O. Box 1065, 41080 Sevilla, Spain.}}

\maketitle 
\begin{abstract}  
{ The interaction due to the exchange of massless neutrinos between neutrons is a
 long-range force. 
Border effects  on this multibody exchange inside a
dense core are studied and computed analytically in $1\ +\ 1$ dimensions. 
We demonstrate in this work that a proper treatment of the star's 
border effect automatically incorporates the condensate contribution 
as a consequence of the appropriate boundary conditions for the neutrino 
Feynman propagator inside the star.
 The total multibody exchange contribution is infrared-safe and  vanishes
 exactly in $1\ +\ 1$ dimensions. The general conclusion of this work is that 
  the border effect does not modify the result that neutrino exchange is
   infrared-safe.
This toy model prepares the ground and gives the tools for the study of 
the realistic $3\ +\  1$ star. }
    \end{abstract}
\vskip -0.3cm
\leftline{} 
\leftline{December 1997}  
\newpage





\renewcommand{\baselinestretch}{1.5}

    The possible connection between the multibody neutrino 
    exchange and the stability of compact stellar objects 
    has recently motivated several relevant works \cite{Fisc96}--\cite{Mor97}.
     The interest raised by Fischbach's original idea \cite{Fisc96} 
     is based on the presumably catastrophic consequences for the
      self-energy of compact objects, such as neutron stars, originated
       by the long-range neutron correlation due to massless neutrino 
       exchange. This catastrophic effect is then invoked to justify the 
       introduction of
        a lower bound for the neutrino mass. In a previous work \cite{Abad96}, 
        we have shown that the total contribution  of the many-body  
    massless neutrino exchanges may be directly and rather easily computed, using 
    an effective Lagrangian,  and results in an  infrared well-behaved star 
    self-energy. 
    The catastrophic result of the
          resummation method is simply due, in our opinion, to the
          fact that it is done outside the radius of convergence of the 
	  perturbative series.

 Smirnov and Vissani 
        \cite{Smir96}, following Fischbach's approach of summing up many-body
        exchange, order by order, showed that the 2-body potential is
         damped by the blocking effects of the neutrino sea \cite{Loeb90} and 
	 conjectured that a
          similar effect for a many-body potential would reduce Fischbach's
           catastrophic effect.
          
 The effect of such a condensate has also been  
 incorporated in our non-perturbative method by using a neutrino 
Feynman propagator inside a dense stellar medium with a 
          condensate term \cite{Abad96}. However, this condensate  does not
           bring any major modification to our conclusion that the total result
	    of the multi-body
            massless neutrino exchange is infrared well-behaved.
          
The objection to our previous work \cite{SmiPP} is that
 we had worked in the approximation where the presence 
 of the neutron star border was negligible. 
 In fact, we stressed in \cite{Abad96} our belief that the neutrino 
 condensate had to
           be understood as a manifestation of the star's   
           border, and a preliminary proof of that statement can be found 
           in ref. \cite{Mor97}. 
           
           In this paper, we shall demonstrate that a proper 
            treatment of the star's border effect automatically incorporates  
            the condensate contribution. And that this is a consequence of 
	    the appropriate
            boundary conditions for the Feynman propagator of the 
  massless neutrino in the interacting medium. 
      Finally, we will show that the border effect does not modify our main result
   that neutrino exchange is infrared-safe.

Our tool in this proof is the computation of the Feynman propagator for the 
massless
 neutrino 
  in the neutron star medium. 
  In \cite{Abad96}, ignoring the border effect and hence 
  using the translational 
  symmetry, we have found the following effective propagator 
\beq
{i\over \rlap/q-b\gamma^0} \ \ ,
\label{1}
\eeq
 where $b\gamma^0$
   accounts for   $Z^0$-exchange diagrams between the neutrino and 
  the neutrons of 
  the medium in the static limit.

The latter is an infrared regular propagator, which can be 
written as follows:

\beq
{i\over \rlap/q-b\gamma^0} = { i \rlap/q_>\over q_>^2+i s\varepsilon} \ \ ,
\label{2}
\eeq

\noindent where $(q_>)^{\mu}=(q_0-b,\vec{q})$. The term 
$i s \varepsilon$ is the infrared regulator, where $s$ is a sign 
that stands for the appropriate ``{\it time convention}". The determination 
of $s$ needs some care: the propagator has been rewritten in terms of the
four-momentum $q_>$, its ``time" component being $(q_0-b)$ instead of  
$q_0$. As we will show, the appropriate boundary conditions, i.e. the ``{\it time
convention"}, should be imposed to keep the usual distribution in the 
$q_0$ complex 
 plane
of the poles for the Feynman propagator: since neutrino (antineutrino) states
correspond to the positive energy (the hole of negative energy) 
solutions, then the positive (negative) energy propagates forward (backward)
in time. This rule implies 
\bea
s=\mbox{sign}(q_0)\mbox{sign}(q_0-b)\ \ .\label{sign}\eea 
It is crucial to insist that by energy we here  meanstrictly $q_0$ and  
not the combination $(q_0-b)$, which expresses the distance of the energy level to 
the bottom of the potential $V\equiv b$. As a consequence, 
$b$ being negative\footnote{In \cite{Abad96}, eq. (2), 
$\omega=|\vec q| \pm b$ \cite{D'Ol92}, is misleading.
 In fact, in an infinite star, the choice between applying
 Feynman's prescription to $q_0$ or $(q_0 -b)$ might appear as open, but when boundary
 conditions are properly taken into account, the choice of $q_0$ becomes
 mandatory.}, we get these dispersion relations for neutrinos and antineutrinos: 
\bea \begin{array}{ll}
E_{\nu} = q_0 =|\vec q| +b \ \ \ \quad\quad \ \ \ \ \mbox{for}\  |\vec q| > |b|\\
E_{\overline{\nu}}=-q_0= \left\{ \begin{array}{ll}
|\vec q| - b  \ \  & \quad \forall |\vec q|\ \ \  \mbox{or}\ ,  \\
-|\vec q| - b \  &\ \ \  \mbox{for} \ |\vec q| < |b|
\end{array} \right.\end{array} 
\label{2b}
\eea

\begin{figure}
\begin{center}
\mbox{\epsfig{file=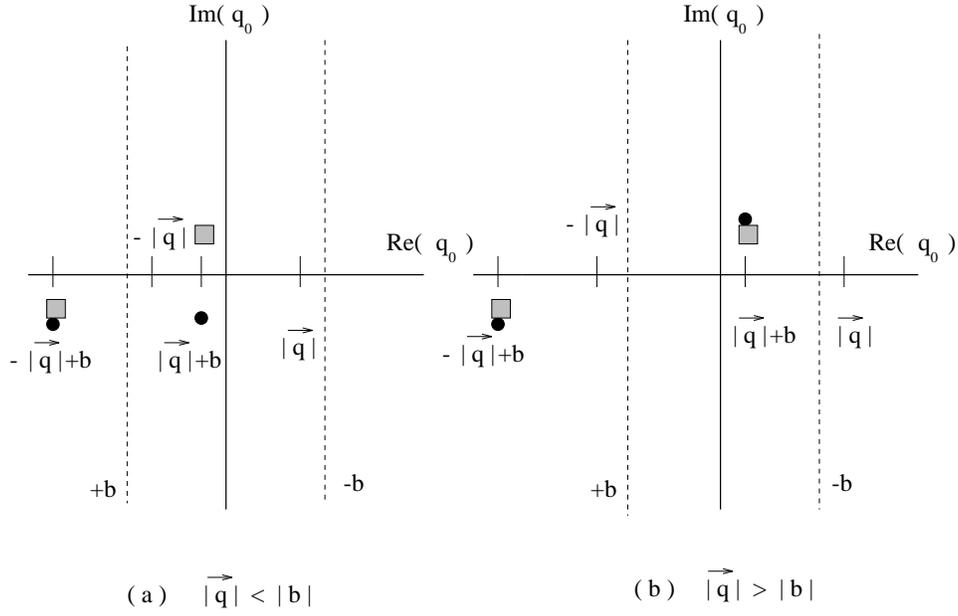,height=8cm}}
\end{center}
\caption{{\small Schematic distribution of the poles of the propagator. The 
black dots and the grey squares represent the poles placed by following the 
time convention $is\varepsilon$ and $+i\varepsilon$, 
 respectively.}}
\label{pole1}
\end{figure}

Finally, the poles are in the first and/or third quadrants in $q_0$ complex 
plane, 
 as shown in fig. \ref{pole1}. Had we taken $s=1$, for a range of values of
  $|\vec q|< |b|$ poles  would have appeared in
 the other quadrants (see fig. \ref{pole1}). Such a pole should be taken into
  account 
when rotating to  Euclidean time. This has not been done correctly 
 in \cite{Abad96}, inducing a minor error which will be discussed below.

 Let us now return  to the central issue of this note, which is the estimation  
 of
 the border effect. As an attempt to have a simple analytic result, 
 we start, in this work,  
 our study with the $1\ + \ 1$ dimensions toy model for the following reasons: 
 i) we will be able to work analytically to the end, ii) the subtle problem 
 of the $i\varepsilon$ convention is easier to master, \\
 iii) thanks to the relative
 simplicity of the problem, we will be able to use different complementary
 techniques and to get a deeper understanding of the physics, which will prove
  most helpful for a more realistic problem,
   i.e. $3\ + \ 1 $ dimensions \cite{border}.
   
   For simplicity, we consider a sharp border located at $z=0$ and use
   an effective neutrino Lagrangian that 
       summarizes the interaction with the neutrons \cite{Mor97}:

\beq
{\cal L}_{eff}=i{\overline{\nu}} (x) \rlap/ \partial  \nu (x) - b \theta(z)
{\overline{\nu}}
 (x) \gamma^0
\nu (x) \ \  ,      \label{lagrangian}
\eeq

\noindent where $b \sim -\sqrt{2} G_F {n_n\over 2} \sim -0.2 \times
10^{-7} {\mbox{GeV}}$, $n_n$ being the neutron density of the star \cite{Abad96}.

The usual definition of the propagator is the following:
\bea
S_F(x, y)=\langle 0 |T\left( \Psi(x) \overline{\Psi}(y)\right) | 0 \rangle \ \
. 
\label{1-3}
\eea
\noindent It is worth stressing at this point  that the time-ordered  product will 
give us the Feynman boundary conditions.

Following Gavela {\it
et al.} \cite{Gav94}, $\Psi(x)$, the general Dirac solution of the problem, 
can be written as:

\bea
\Psi(x)=\sum_{\pm\vec{k}, h} b_{h}(\pm\vec{k})\nu^h_{\pm\vec{k}}(\vec{x})
e^{-iEt} \ + \ d^{\dag}_{h}(\pm\vec{k})\nu^{A\ h}_{\pm\vec{k}}(\vec{x})
e^{iEt} \ \ , 
\label{2-3}
\eea
\noindent where $\nu$ are the eigenstates of the Dirac
 Hamiltonian derived from the
Lagrangian (\ref{lagrangian}). 
In the second quantization, the coefficients $b$ and $d^{\dag}$ become annihilation and 
creation operators, respectively.

The functions $\nu$ can be obtained from the following complete set of solutions:
\bea
&&\nu_{\pm{\vec k}}^{h}(z,t)= e^{-iE(t\mp z)}u_{h}(\pm{\vec k})\left\{
\theta(-z)+e^{\mp ibz}\theta(z)\right\} \ \ , \nn \\
&&\nu_{\pm {\vec k}}^{A \ h}(z,t)= e^{iE(t\mp z)}v_{h}({\pm\vec k})\left\{
\theta(-z)+e^{\mp ibz}\theta(z)\right\} \ \  \ ,
\label{4-4}
\eea
\noindent which are obtained by solving  the equation of motion for the $1\ 
+\  1$ 
 Lagrangian of 
 eq. (\ref{lagrangian}),
\bea
\left\{ \gamma^0\left(i\partial_t-b\theta(z)\right)+i \gamma^1 
\partial_1\right\}\nu({\vec x},t) = 0 \ \ .
\label{1-2b}
\eea
 
In eq. (\ref{4-4}), the index $h$ expresses the chirality of the solution
and $A$ denotes the antiparticles. Note that the chirality is the same on both
sides of the border, as expected from the chirality-conserving Lagrangian of 
eq. (\ref{lagrangian}). 

In the region $z>0$ (inside the star), neutrinos have $k_z=\pm(E-b)$ and antineutrinos 
 have $k_z=\pm(E+b)$, while in the region $z<0$ (outside the star), they have both 
 $k_z=\pm E$. The subscripts $\pm {\vec k}$ depend on the choice of the sign for 
$k_z$. In Eq. (\ref{4-4}) $u$ and $v$ are the Dirac spinors for
particles and antiparticles, respectively,

\bea
\rlap/k_{>,<}^{\pm}\left\{ \begin{array}{r} u_{h}(\pm \vec{k}) \\ 
v_{h}(\pm\vec{k}) \end{array} \right\} \  =\ 0  \ ,
\label{4-4b}
\eea

\noindent where $k_>^{\pm}=(E - b, \pm (E -
b) )$ for
neutrinos, $k_>^{\pm}=(E + b, \pm (E + b) )$ for antineutrinos  
and $k_<^{\pm}=(E, \pm E)$ for both.

Equations (\ref{2-3}) and (\ref{4-4}) imply a definite choice of the zero 
energy level that  corresponds to the standard choice of the free neutrinos
 far outside the star, $z\to -\infty$, clearly the only admissible choice.
 Physically this choice is related to the use of a stationary Lagrangian, 
 (\ref{lagrangian}), which means that we assume that the system has relaxed to
 equilibrium, implying plane-wave solutions that extend outside and 
 inside the star.
 These plane waves ``know'' the zero energy level from their outer domain.
 This choice, combined with the time-ordered
 product of  eq. (\ref{1-3}), implies the $+is\varepsilon$ time regulator in
  momentum 
 space as announced before.
 
In momentum space, the propagator can be written as 
\bea
\begin{array}{l}
S_F(q^f,q^i)\gamma^0={i\over 2\pi}\int_0^{\infty}dk_0\left[
\widetilde{\nu_{\pm\vec{k}}^{h}(q_z^f)}
\left(\widetilde{\nu_{\pm\vec{k}}^{h}(q_z^i)}\right)^{\dag}
{1\over q_0-k_0+i\varepsilon} \right.  \\
\left. + \ \widetilde{\nu_{\pm\vec{k}}^{A \ h}(q_f^f)}
\left(\widetilde{\nu_{\pm\vec{k}}^{A \ h}(q_z^i)}\right)^{\dag}
{1\over q_0+k_0-i\varepsilon} \right] \ \ \ ,\end{array}
\label{4-7}
\eea
\noindent where $q^{f,i}=(q_0, q_z^{f,i})$. 
Notice that the character of creation or annihilation for the 
operators in eq. (\ref{2-3}) is fixed by taking the matter-free vacuum 
as reference, i.e. it is a consequence of our choice of the zero 
energy level.  
$\widetilde{\nu}$ are the Fourier transform of the eigenstates 
$\widetilde{\nu(q_z)}=\int_{-\infty}^{+\infty}dz\  e^{-iq_z z}\nu(z)$:

\bea
\widetilde{\nu_{\pm{\vec k}}^{h}(q_z)}=
i\left( {1\over q_z\mp k_0+i\varepsilon}-{1\over
q_z\mp (k_0-b)-i\varepsilon}\right)u_{h}(\mp{\vec k}) \ \ , \nn \\
\widetilde{\nu_{\pm{\vec k}}^{A \ h}(q_z)}=
i\left( {1\over q_z\pm k_0+i\varepsilon}-{1\over
q_z\pm(k_0+b)-i\varepsilon}\right)v_{h}(\pm{\vec k}) \ \ .
\label{4-6}
\eea

\noindent We took  $E=k_0$ in order to make the notation uniform.
Applying this last result in eq. (\ref{4-7}), with the appropriate 
analytic continuation of the integration variable $k_0$, and by following 
the adequate integration contour (see appendix \ref{details}), we obtain for 
the $1\ +\ 1$ propagator:

\bea
\begin{array}{l}S_F(q^f,q^i)= 2\pi\delta\left( q_z^f-q_z^i \right) 
{i\over (\rlap/q_>)^*}  
\ + \ {b\over 2}\ {1\over q_z^f -q_z^i+i\varepsilon} \\
\times \left\{ {1\over
(\rlap/q_>^i)^*}\gamma^0{1\over \rlap/q_<^f} 
\left(1+ \mbox{sign}(q_0)\alpha_z \right) \right.  
\left. + {1\over (\rlap/q_>^f)^*}\gamma^0{1\over \rlap/q_<^i}
\left(1-\mbox{sign}(q_0)\alpha_z \right)
\right\} \ \ , \end{array}
\label{4-11}
\eea

\noindent where we define:

\bea 
&&{1\over \rlap/q_<}={1\over \rlap/q}= {\rlap/q_< \over q_<^2+i\varepsilon} 
\ \mbox{with} \ \ \left(q_<\right)^{\mu}=q^{\mu}=(q_0, {\vec q}) \ \ , \nn
\\ 
&&{1\over (\rlap/q_>)^*}={\rlap/q_> \over q_>^2+is\varepsilon} 
\ \mbox{with} \ \ \left(q_>\right)^{\mu}=(q_0-b, {\vec q}) \ \ ,
\label{4-12}
\eea

\noindent with $s=\mbox{sign}(q_0)\mbox{sign}(q_0-b)$   
and $\alpha_z=\gamma^0\gamma^1$. 
The location of the poles in the complex plane have already been depicted 
in fig. \ref{pole1}. 

Anticipating over the vacuum energy calculation, it is interesting to notice 
that the second term in the r.h.s of eq. (\ref{4-11}) does not contribute
 when the 
border is sent to infinity; one is then left with only the first term, which 
is exactly the one of eq. (\ref{2}) for the infinite star with the 
same good regularization.
 To convince ourselves, eq. (\ref{4-11}) can be rewritten as follows: 

\bea
\begin{array}{l}S_F(q^f,q^i)= 2\pi i \delta\left( q_z^f-q_z^i \right)\left\{   
{1\over (\rlap/q_>)^*} - {b\over 2}{1\over
(\rlap/q_>)^*}\gamma^0{1\over \rlap/q_<} \right\} \\
+{b\over 2}{\cal P}\left( {1\over q_z^f -q_z^i} \right) 
\left\{ {1\over
(\rlap/q_>^i)^*}\gamma^0{1\over \rlap/q_<^f} 
 - {1\over (\rlap/q_>^f)^*}\gamma^0{1\over \rlap/q_<^i}
\right\}\mbox{sign}(q_0)\alpha_z \ \ ,\end{array} 
\label{4-11b}
\eea

\noindent where ${\cal P}$ stands for the principal value. For an observer
located at $z \to \infty$, which means that he does not see the star 
border, restoring the translational symmetry, i.e.  $q_z^f \to
q_z^i$ (when $z \to \infty$), is a good
approximation. Consequently, the second term in the r.h.s of eq. (\ref{4-11b}) 
will not contribute when $z \to
\infty$ \footnote{In fact, we can see in 
eqs. (\ref{01-33}) and (\ref{01-34}) below that the break-up of the
translational symmetry is expressed in 
$e^{i(q_z^f-q_z^i)z}$, which gives the dependence on $z$. This
oscillating  term, in the
limit $z \to \infty$, will destroy any contribution to the integral
coming from the part of the principal value of the
propagator.}.  
Furthermore, it can be seen from eq. (\ref{01-33}) below that after
 the integration
over the momenta, as suggested by Schwinger's method to obtain the
vacuum energy, the second term  in the 
r.h.s of eq. (\ref{4-11}) does not  contribute 
to the energy density $w(z\to +\infty)$.

The expression for $1/(\rlap/q)^*$, given by eq. (\ref{4-12}), can be
appropriately rewritten as 
\bea
{i\over (\rlap/q_>)^*} = i\left\{ {1\over \rlap/q_>}+2\pi i
\rlap/q_>\delta\left(q_>^2\right)\theta(q_0)
\theta(b-q_0)  \right\} \ \ .
\label{4-14}
\eea

This last equation is a concrete demonstration that we have,
 in $1\  +\  1$ dimensions, generated the condensate contribution in a natural way 
 to all orders. In refs. \cite{Abad96} and \cite{Smir96}, the same condensate
  term in the r.h.s of eq. (\ref{4-14}) was introduced by hand. We should stress that
in ref. \cite{Abad96} we  initially  computed the weak energy density by
using the $+i\varepsilon$ convention (first term of eq. (\ref{4-14})),
 but we did not take into account the pole $q_0=|\vec{q}|+b$, in the
second quadrant for $|\vec{q}|<|b|$ (see fig. \ref{pole1}) when
rotating to Euclidean 
time. This explains our non-zero result for the energy density
when we have fixed the good boundary conditions by introducing the second
term of eq. (\ref{4-11}) for the propagator.

The existence of a certain neutrino condensate due to the
interaction of neutrinos and the stellar matter background was initially
 proposed
 by Loeb \cite{Loeb90}: neutrinos are trapped inside the star and antineutrinos 
 are repelled. The relevant
physics concerning the neutrino propagator in the scenario we
described above would appear by introducing the appropriate boundary
conditions. Equations (\ref{4-12}) and (\ref{4-14}) give a confirmation of the
idea proposed in ref. \cite{Abad96}: the condensate is a consequence of 
the existence of a border.

This condensate is physically understandable. As we have tuned the level of the 
Dirac
 sea outside the star (to the left) and as our states extend over all space, far 
 inside the star (to the right), the level corresponds to filling a Fermi sea 
 above the bottom of the potential $b<q_0<0$. 
This  obviously induces a Pauli blocking effect\footnote{Smirnov and Vissani
 \cite{Smir96} have proved that the condensate term,
  obtained above in a natural way, generates the damping of the 
  2-body neutrino-exchange potential.}. Equation (\ref{2b}) anticipates this result: 
  $|b|$ is a lower bound for the momentum of the $q_0$-positive states.

With these tools, let us compute the total neutrino-exchange contribution to
 the 
energy of the star, which is nothing else than the difference between the vacuum
 energy of our effective
 theory eq. (\ref{lagrangian}) and 
the vacuum energy of free neutrinos (without the star). 
Following Schwinger's method \cite{shwinger}, the vacuum energy in the presence
 of an external field is 
 given by tracing the Hamiltonian multiplied by the propagator, as done in 
 refs. \cite{Fisc96} and \cite{Abad96}.
Now, we compute the time-independent vacuum energy  density
\bea
\begin{array}{l}
 w(z)\equiv -{i} \partial_t \mbox{tr} 
[\gamma^0 \left(S_F(z,t;z',t)-S^{(0)}_F(z,t;z',t)\right)P_L ]_{z'\to z} 
\\ =
 \int {dq_z^f \ dq_z^i \ dq_0 \over
(2\pi)^3} \ 
\mbox{tr}\left[q_0\gamma^0 e^{i(q_z^f-q_z^i)z}
\left(S_F(q^f,q^i)-S_F^{(0)}(q^f,q^i)\ 2\pi
\delta\left(q_z^f-q_z^i\right)\right)P_L \right] \ \ ,  \end{array}
\label{01-33} 
\eea

\noindent where $S^{(0)}_F$ is the free neutrino propagator and 
$P_L={1-\gamma_5\over 2}$.   
We should remark that the $e^{i(q_z^f-q_z^i)z}$ in the second line of 
this equation is generated by the break-up of the translational 
 symmetry due to the border.

 Now, by applying the expression of eq. (\ref{4-11}) for the 
propagator in this last equation, we obtain:

\bea
&&w(z) \ =  - \int \ {dq_z \ dq_0 \over
(2\pi)^2} \mbox{tr}\left[q_0\gamma^0 
\left( {i\over (\rlap/q_>)^*} - {i\over \rlap/q_<} \right) P_L
\right] \nn  \\
&&- {b\over 2}\int \ {dq_z^f \ dq_z^i
 \ dq_0 \over (2\pi)^3} \ {1\over q_z^f -q_z^i+i\varepsilon}  \times  \nn \\
&&\mbox{tr}\left[ e^{i(q_z^f-q_z^i)z}\left\{ {1\over
(\rlap/q_>^i)^*}\gamma^0{1\over \rlap/q_<^f} 
\left(1+\mbox{sign}(q_0)\alpha_z \right) \right. \right. 
\left. \left. + {1\over (\rlap/q_>^f)^*}\gamma^0{1\over \rlap/q_<^i}
\left(1-\mbox{sign}(q_0)\alpha_z \right)
\right\}P_L\right]\ \ .
\label{01-34}
\eea

\noindent It should be noted that the first term of the r.h.s can be
 appropriately rewritten by using the Schwinger--Dyson relation, which has
 been shown in ref. \cite{Abad96} for the effective propagator (\ref{1})
in the case of the infinite star \footnote{Of course the Schwinger--Dyson
relation is also valid for the full propagator, including the border (13),
but it is  rather complicated to write it explicitly, and serves no great purpose here.}.
 However, the calculation is rather cumbersome, even if it presents 
no special difficulty and leads to the amazing  result:

\beq 
w(z)=0 \ .
\eeq

As surprising  as this result  may seem, it can be understood physically in a
 rather simple way.
 One must remember \cite{shwinger} that $w(z)$ is nothing else than:

\bea
\sum\!\!\!\!\!\!\!\int _{n_-} E_{n_-} \Psi^{\dag}_{n_-}(z) \Psi_{n_-}(z)
-\sum\!\!\!\!\!\!\!\int _{n_-}E^{(0)}_{n_-} \Psi^{\dag(0)}_{n_-}(z) \Psi^{(0)}
_{n_-}(z)\ \ ,
\label{sumE-}
\eea

\noindent where  $n_-$ labels the negative energy states and $(0)$
 refers to the matter-free vacuum.

These integrals are ultraviolet-divergent and we will allow the interchange
 of the sum and the
 integral as a regularization method.
This was implicitly done in eq. (\ref{01-34}). 
Looking now  at the solutions of eq. (\ref{4-4}) of the Hamiltonian with and  
without ($b=0$) the star, we  first that remark there is a one-to-one 
correspondance of 
the states in the two situations and with the same energy; furthermore 
the probability density of the states are all equal to 1 for all $z$.
 More explicitly, the border does not disturb the wave functions, except for 
 a phase. Consequently, after regularization (interchanging the sum and the 
 difference)
  each term  in the sum of eq. (\ref{sumE-})
 vanishes.

The ($1 \ +\ 1$)-dimensional problem presents the particularity that,
 for a given wave-plane solution, 
 the sign of the momentum determines a positive correspondence
  between $z= \ + \ (-) \infty$ and $t= \ + \ (-) \infty$, or vice versa. 
  Massless fermions are not reflected by a one-dimensional potential.
   However, the potential introduced through the Lagrangian
(\ref{lagrangian}) does not allow us to identify, for a given process,
 asymptotically free states for $t=\pm \infty$, because there is no 
  ``vacuum'' to
 the right,
 and hence the S-matrix formalism 
 is not adequate in a world defined by such a Lagrangian. 
 In order to avoid the latter, we can take a second border. 
 In this case, one can easily see that the phase shift taken by 
 the neutrino states keeps the S-matrix diagonal: the star  is ``transparent" 
  for the neutrino propagation. Therefore, by applying eq. (\ref{sumE-}), 
  we can now conclude that the diagonality of the S-matrix justifies the null 
  result for the weak energy density 
  \footnote{Despite the major analytic complexity introduced
   by the second border, we have computed the Feynman propagator 
   and verified, by following the first method presented in this work,
    that the weak energy-density result is exactly zero.}.

The generalization to the ($3\ +\ 1$)-dimensional problem is
  difficult even if the border is simplified to a flat one because of the 
  presence of a refraction index, of a consequent modification of the probability 
 density of the waves at the border, including non-penetrating or non-outgoing
 waves. This ($3\ +\  1$)-dimensional problem 
  will be analytically and numerically studied in a 
  forthcoming work \cite{border}. 
  
However, two main conclusions of this paper will remain valid 
in $3\ +\ 1$ dimensions:\\
 i) the natural connection between the neutrino sea and the border, 
  ii) the proper definition of the $is\epsilon$ infrared
regulator for the propagator, or, equivalently, the correct definition of the
zero energy level of the Dirac sea. On the other hand, the vanishing of the
energy density $w(z)$ will not remain valid in $3\ +\ 1$ dimensions.

Finally, this work is a confirmation of 
the conclusion of ref. \cite{Abad96}: after correcting the mistake 
of forgetting the pole in the analytic continuation, 
and adding the condensate, which is proved here to be directly connected to 
the border,
 the multibody exchange of massless neutrinos results in an infrared-well
 -behaved contribution for the neutron star self-energy.

\bigskip
\bigskip

{\large \bf Acknowledgements}

This 
work has been partially supported by Spanish CICYT, project 
PB 95-0533-A. We are specially grateful to B. M. Gavela for the discussions
that initiated this work. 
A. Abada thanks A. Smirnov for inspiring discussion. 
J. Rodr\'{\i}guez--Quintero thanks M. Lozano for many important and helpful 
discussions and comments.

\newpage

\appendix
\centerline{\bf Details of integration in $1\ +\ 1$ dimensions}\label{details}

To compute the Feynman neutrino propagator in the momentum space, we 
should apply the results for the Fourier-transformed eigenstates
 (\ref{4-6}) in eq. (\ref{4-7}). After some tedious transformations, we obtain:

\bea
{i\over 2\pi}\int_0^{+\infty}dk_0\left(I_1+I_2+I_3+I_4\right) \ \ ,
\label{C-8}
\eea

\noindent where

\bea\begin{array}{l}
I_1=
{1\over {1\over
b}\left(q_z^f-k_0\right)\left(q_z^f-k_0+b\right)+i\varepsilon} \times
 \\   \\
{1\over {1\over
b}\left(q_z^i-k_0\right)\left(q_z^i-k_0+b\right)-i\varepsilon}   
\ {1\over 2}\left(\gamma^0-\gamma^1\right)\gamma^0{1\over
q_0-k_0+i\varepsilon} \ \ , \\ \\
I_2=
{1\over -{1\over
b}\left(q_z^f+k_0\right)\left(q_z^f+k_0-b\right)+i\varepsilon}  \times
 \\ \\
{1\over -{1\over
b}\left(q_z^i+k_0\right)\left(q_z^i+k_0-b\right)-i\varepsilon}   
\ {1\over 2}\left(\gamma^0+\gamma^1\right)\gamma^0{1\over
q_0-k_0+i\varepsilon} \ \ ,  \\ \\
I_3=
{1\over {1\over
b}\left(q_z^f+k_0\right)\left(q_z^f+k_0+b\right)+i\varepsilon}  \times
\\ \\
{1\over {1\over
b}\left(q_z^i+k_0\right)\left(q_z^i+k_0+b\right)-i\varepsilon}   
\ {1\over 2}\left(\gamma^0-\gamma^1\right)\gamma^0{1\over
q_0+k_0-i\varepsilon} \ \ ,\\ \\
I_4=
{1\over -{1\over
b}\left(q_z^f-k_0\right)\left(q_z^f-k_0-b\right)+i\varepsilon}  \times
\\ \\
{1\over -{1\over
b}\left(q_z^i-k_0\right)\left(q_z^i-k_0-b\right)-i\varepsilon}   
\ {1\over 2}\left(\gamma^0+\gamma^1\right)\gamma^0{1\over
q_0+k_0-i\varepsilon} \ \ .
\label{4-9}\end{array}
\eea

\begin{figure}[hbt]
\begin{center}
\mbox{\epsfig{file=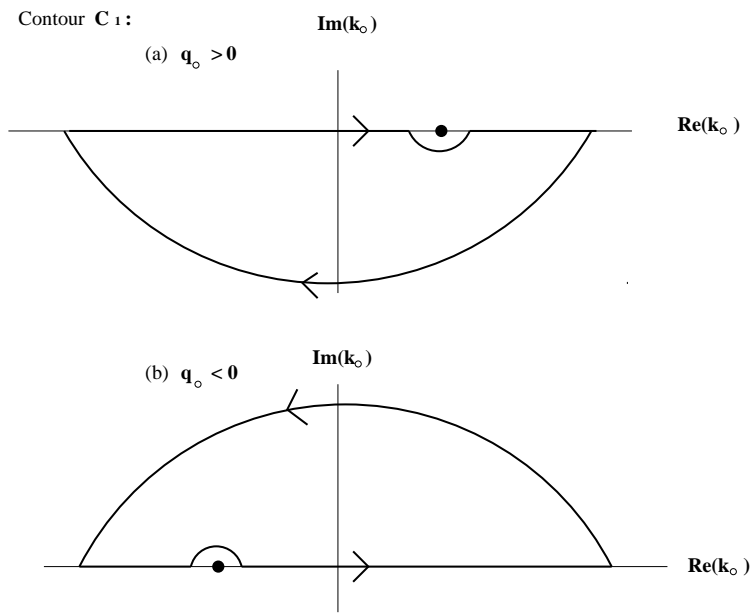,height=8cm}}
\end{center}
\caption{{\small Integration contour $C_1$. (a) 
$q_0>0$ case; (b) $q_0<0$ case.}}
\label{f-1}
\end{figure}

\noindent By taking into account the form of the different expressions
for $I_i$ in eqs. (\ref{4-9}), we have:

\bea
\int_0^{\infty}dk_0 \left( I_1+I_3\right) = \int_{C_1}dk_0 I_1 \ \ ,
\nn \\
\int_0^{\infty}dk_0 \left( I_2+I_4\right) = \int_{C_1}dk_0 I_2 \ \ ,
\label{4-10}
\eea

\noindent where the integration in the r.h.s should be done in the complex plane by following the
contour $C_1$ (see fig. \ref{f-1}).

Identifying the relevant poles for each case and applying Cauchy's theorem, we obtain the $1\ +\ 1$ propagator given 
above by eq. (\ref{4-11}).

\end{document}